\documentclass[twocolumn,showpacs,pre]{revtex4}
\usepackage{amsfonts}
\usepackage{graphicx}
\usepackage{subfigure}

\input{tcilatex}

\begin{document}

\title{Moving embedded lattice solitons }
\author{B. A. Malomed$^{1}$, J. Fujioka$^{2,\ast }$, A. Espinosa-Cer\'{o}n$%
^{2}$, R. F. Rodr\'{\i}guez$^{2,\ast }$ and S. Gonz\'{a}lez$^{2}$}
\affiliation{$^1$ Department of Interdisciplinary Studies, Faculty of Engineering, Tel
Aviv University, Tel Aviv 69978, Israel\\
$^{2}$Instituto de F\'{\i}sica, Universidad Nacional Aut\'{o}noma de M\'{e}%
xico, Apdo. Postal 20-364, M\'{e}xico D.F. 01000, M\'{e}xico.\\
$^{\ast }${\small Fellows of SNI, M\'{e}xico.}}

\begin{abstract}
It was recently proved that solitons \textit{embedded }in the spectrum of
linear waves may exist in discrete systems, and explicit solutions for
isolated unstable embedded lattice solitons (ELS) of a
differential-difference version of a higher-order NLS equation were found
[Physica D \textbf{197} (2004) 86]. The discovery of these ELS gives rise to
relevant questions such as the following: are there \textit{continuous
families} of ELS?, can ELS be \textit{stable}?, is it possible for ELS to
move along the lattice?, how do ELS interact?. The present work addresses
these questions by showing that a novel equation (a discrete version of a
complex modified KdV equation which includes next-nearest-neighbor
couplings) has a two-parameter continuous\ family of exact ELS. These
solitons can move with arbitrary velocities across the lattice, and the
numerical simulations demonstrate that these ELS are completely stable.
Moreover, the numerical tests show that these ELS are robust enough to
withstand collisions, and the result of a collision is only a shift in the
positions of the solitons. The model may apply to the description of a
Bose-Einstein condensate with dipole-dipole interactions between the atoms,
trapped in a deep optical-lattice potential.

%%\noindent Key words: embedded solitons, lattice solitons, discrete systems,
%%finite-difference equations, modified Korteweg-de Vries equation,
%%Ablowitz-Ladik equation
\end{abstract}

\maketitle

%\baselineskip 1.0 cm

%\pagebreak

\textbf{In nonlinear systems where solitons can exist, the propagation of
small-amplitude linear waves, which obey the linearized version of the
nonlinear equations, is possible too. \ However, for a soliton to exist, it
is absolutely necessary that no resonances occur between the soliton and
these linear waves. Otherwise, the soliton would decay due to an energy
transfer towards the linear waves. Based on this no-resonance argument, it
was frequently assumed that the solitons' internal frequencies could not be
contained in the linear spectrum of the system, \textit{i.e}., they could
not lie within the band of frequencies permitted to linear waves. However,
at the end of the nineties exceptions to this rule were found, and a special
type of solitons were discovered, which do not resonate with linear waves,
in spite of having frequencies immersed in the spectrum of these waves. In
1999 these peculiar solitary waves were given the name of \textit{embedded
solitons}} \textbf{(ES), and} \textbf{in the following years a number of
models supporting ES were identified. Most of these models describe
continuous systems. However, some examples of discrete ES were recently
found too. These \textit{embedded lattice solitons} (ELS) are isolated
solutions which are stable against small perturbations in the linear
approximation, but are nonlinearly unstable. The discovery of these isolated
unstable ELS triggered the search for models admitting \textit{continuous
families }of \textit{stable} ELS. In this article we present a novel
differential-difference equation which has a two-parameter \textit{%
continuous family} of exact ELS. These solitons can move with arbitrary
velocities across the lattice, and the numerical simulations show that they
are \textit{completely stable} solutions. }

\section{Introduction}

In the late nineties, a new type of solitary waves was identified and given
the name of \textit{embedded solitons }(ES). At first, ES were found in
nonlinear-optical models including quadratic ($\chi ^{(2)}$) nonlinearities 
\cite{Fujioka97}-\cite{Yang2001A}\/, and later they were discovered in
hydrodynamic \cite{Yang2001}\/ and liquid-crystal dynamical models \cite%
{Rodriguez2003}\/. Quite recently, ES were found in discrete systems too, 
\textit{viz}., in a finite-difference version of a higher-order NLS
(nonlinear Schr\"{o}dinger) equation \cite{Silvia}, and in a model of an
array of linearly-coupled waveguides with the $\chi ^{(2)}$ and $\chi ^{(3)}$%
\/\/ (cubic) nonlinearities \cite{Kazu}. The distinctive feature of \ the%
\emph{\ }ES is that their internal frequencies and, in some cases, also
their velocities,\emph{\ }fall in a spectral region occupied by linear wave
modes, where regular solitons cannot exist, as they would decay into
radiation. That is, ES are exceptional states that \textit{do not} emit
linear waves, despite being in resonance with them. A qualitative
explanation to the existence of ES was proposed in Ref. \cite{Yang99} (see
also \cite{PhysD}\/), and more quantitative arguments were provided in Ref. 
\cite{Espinosa}\/.

A number of generic features of the ES found in continuous systems have been
identified. In particular, ES are usually isolated solutions, although 
\textit{continuous families} of ES may exist too, under special symmetry
conditions \cite{Yang2001,Yang2003,continuous-families}. The ES may be
stable against small perturbations in the linear approximation, but general
arguments and direct simulations demonstrate that they are usually unstable
at the second order, which is why they are often called \textquotedblleft
semi-stable" solutions. Nevertheless, examples of completely stable ES have
been also found \cite{Rodriguez2003, Yang2003, Dmitry}, and their stability
has been demonstrated in Refs. \cite{Yang2003, Dmitry}.

Much less is known about embedded lattice solitons (ELS) in discrete
systems. Only a few examples of models admitting isolated nonlinearly
unstable ELS have been found \cite{Silvia,Kazu}. Some issues which remain to
be analyzed are, for example, the possibility of the existence of \textit{%
continuous families} of ELS, and to determine if they may be completely
stable. Furthermore, a challenging question is whether ELS may move across
the lattice with arbitrary constant velocities. We address these issues in
the present work, introducing a new nonlinear lattice model which gives rise
to exact solutions for ELS which feature all these properties.

The paper is organized as follows. In Sec. 2 we present the new model and
its continuous family of \textit{moving} lattice solitons. In Sec. 3 we show
that these moving solitons are indeed ELS\textit{.} In Sec. 4 the behavior
of perturbed and colliding ELS is investigated by means of direct
simulations which, in particular, demonstrate that more general solitons
with different amplitudes exist too. Section 5 concludes the paper.

\section{The model and its family of moving lattice-soliton solutions}

Motivated by the recent discoveries of continuous families of ES in the
complex modified Korteweg-de Vries (cmKdV) equation \cite{Rodriguez2003} and
in a generalized third-order nonlinear Schr\"{o}dinger equation \cite%
{Yang2003, Dmitry}, in the present paper we will consider a
differential-difference equation of the following form:%
\begin{eqnarray}
\dot{r}_{n} &=&-2\varepsilon _{0}\left( r_{n+1}-r_{n-1}\right) +\varepsilon
_{0}\left( r_{n+2}-r_{n-2}\right)  \nonumber \\
&&+\rho \left\vert r_{n}\right\vert ^{2}\left( r_{n+1}-r_{n-1}\right)
-\delta \left\vert r_{n}\right\vert ^{2}\left( r_{n+2}-r_{n-2}\right) 
\nonumber \\
&&  \label{DDE2}
\end{eqnarray}%
where $r_{n}\left( t\right) $ is a complex variable, the overdot stands for $%
d/dt$, and $\varepsilon _{0}$, $\rho \emph{\ }$and $\delta $ are real
constants. At fixed $t$, the discrete values $r_{n}\left( t\right) $ can be
considered as the values of a continuous function $u\left( x,t\right) $,
evaluated at equally spaced lattice points. In other words, we can consider
that $r_{n}\left( t\right) \equiv u\left( x=n\Delta x,t\right) $, where $%
\Delta x$ is the lattice spacing. If we now consider the limit when $\Delta
x\rightarrow 0$, a standard limiting procedure (described in detail in Ref. 
\cite{Silvia}) shows that the Eq. (\ref{DDE2}) goes over into the equation: 
\begin{equation}
\frac{\partial u}{\partial t}=\varepsilon \frac{\partial ^{3}u}{\partial
x^{3}}+\frac{1}{6}\left( \rho -2\delta \right) \gamma \left\vert
u\right\vert ^{2}\frac{\partial u}{\partial x},  \label{cmKdV2}
\end{equation}%
where $\varepsilon =2\varepsilon _{0}\left( \Delta x\right) ^{3}$ and $%
\gamma =12\Delta x$. If $\rho -2\delta =6$ Eq. (\ref{cmKdV2}) is precisely
the cmKdV equation studied in \cite{Rodriguez2003}. It is worth observing
that the coefficient in front of the nonlinear term in the cmKdV equation
may be positive or negative, depending on the physical application. For
example, in the description of light propagation in liquid-crystal
waveguides, this coefficient is positive \cite{Rodriguez2003}. On the other
hand, the derivation of the cmKdV equation for strongly dispersive waves in
a weakly nonlinear medium by means of a multiple-scale expansion, presented
in Ref. \cite{Degasperis}, leads to a negative nonlinear coefficient.

If we neglect the nonlinear terms in Eq. (\ref{DDE2}) we arrive at a linear
equation which describes a \textit{dispersive lattice} (\textit{i.e.}, a
discrete version of the linear dispersive equation $u_{t}=\varepsilon
u_{xxx} $). However, for the existence of ELS solutions. both nonlinear
terms in Eq. (\ref{DDE2}) are needed.

By means of the \textit{staggering transformation}, 
\begin{equation}
r_{n}(t)\equiv \left( -i\right) ^{n}R_{n}(t)  \label{staggering}
\end{equation}%
Eq. (\ref{DDE2}) can be cast in an alternative form,%
\[
i\dot{R}_{n}+2\varepsilon _{0}\left( R_{n+1}+R_{n-1}\right) -\rho \left\vert
R_{n}\right\vert ^{2}\left( R_{n+1}+R_{n-1}\right) 
\]%
\begin{equation}
+i\varepsilon _{0}\left( R_{n+2}-R_{n-2}\right) -i\delta \left\vert
R_{n}\right\vert ^{2}\left( R_{n+2}-R_{n-2}\right) =0.  \label{R}
\end{equation}%
In this form, it is an extension of the well-known integrable Ablowitz-Ladik
(AL) model \cite{AL}, that corresponds to the terms in the first row in Eq. (%
\ref{R}). The first (linear) additional term accounts for the \textit{%
higher-order lattice diffraction} \cite{HOD}. As shown in Ref. \cite{HOD},
the third-order lattice diffraction can be induced by biasing a lattice
(with the coupling between next-nearest neighbors) by a phase gradient
applied along it (in particular, by means of the \textit{%
diffraction-management }technique in an array of nonlinear optical
waveguides \cite{Silberberg}). Then, the nonlinear terms in Eq. (\ref{R})
may be regarded as generic nonlinear corrections to the second- and
third-order lattice diffraction. In physical systems, nonlinear terms of
these types should be induced by long-range nonlinear interactions between
sites of the lattice. A realization of this is known in Bose-Einstein
condensates (BECs) with \emph{dipole-dipole interactions} between atoms,
trapped in a deep periodic potential (which is created by means of an
optical lattice) \cite{Liu}. Note that Eq. (\ref{R}) does not include any
on-site nonlinearity, which can be eliminated in the BEC by means of the
known technique of the Feshbach resonance \cite{FR}.

A basic finding of this work is that Eq. (\ref{DDE2}) has a family of exact
soliton solutions, provided that $\varepsilon _{0}\rho <0$ and $\delta
>2\rho $, of the following form: 
\begin{equation}
r_{n}\left( t\right) =A_{\mathrm{sol}}\,\mathrm{sech}\left( \frac{n\Delta
x-at}{w}\right) \,e^{i\,\left( qt\,+\,nQ\Delta x\right) },  \label{soliton}
\end{equation}%
where $Q$ and $\Delta x$ are free parameters, and the values of $A_{sol}$, $%
w $, $a$ and $q$ are determined by the following equations: 
\begin{equation}
A_{\mathrm{sol}}^{2}=\frac{\varepsilon _{0}}{\rho }\left( 2-\frac{\delta }{%
\rho }\right) ,  \label{A}
\end{equation}%
\begin{equation}
\mathit{\cosh }^{2}\left( \frac{\Delta x}{w}\right) =\frac{\delta }{2\rho },
\label{w}
\end{equation}%
\begin{eqnarray}
\frac{a}{\Delta x} &=&-2\varepsilon _{0}\mathit{\cos }\left( 2Q\Delta
x\right) \frac{w}{\Delta x}\mathit{\sinh }\left( \frac{2\Delta x}{w}\right) 
\nonumber \\
&&+4\varepsilon _{0}\mathit{\cos }\left( Q\Delta x\right) \,\frac{w}{\Delta x%
}\mathit{\sinh }\left( \frac{\Delta x}{w}\right) ,  \label{a}
\end{eqnarray}%
\begin{eqnarray}
q &=&2\varepsilon _{0}\mathit{\sin }\left( 2Q\Delta x\right) \mathit{\cosh }%
\left( \frac{2\Delta x}{w}\right)  \nonumber \\
&&-4\varepsilon _{0}\mathit{\sin }\left( Q\Delta x\right) \mathit{\cosh }%
\left( \frac{\Delta x}{w}\right) .  \label{q}
\end{eqnarray}%
For fixed values of the coefficients $\varepsilon _{0}$, $\rho $ and $\delta 
$, the Eqs. (\ref{soliton})-(\ref{q}) define a continuous family of \textit{%
moving} lattice solitons, all with the same amplitude, due to Eq. (\ref{A}),
but different widths $w$ and velocities $V=a/\Delta x$ (the latter being
defined with respect to the discrete coordinate $n$). Because the value of $%
\Delta x/w$ is fixed by Eq. (\ref{w}), and $\left\vert \cos \left( 2Q\Delta
x\right) \right\vert \leq 1$, Eq. (\ref{a}) shows that the velocity always
takes values within a limited interval, 
\begin{equation}
\left\vert \frac{a}{\Delta x}\right\vert \leq 2\left\vert \varepsilon
_{0}\right\vert \frac{w}{\Delta x}\mathit{\sinh }\left( \frac{2\Delta x}{w}%
\right) .  \label{interval}
\end{equation}

It is relevant to stress that, contrary to a frequently made assumption,
explicit counter-examples \cite{Flach, Kevrekidis2003}\ demonstrate that the
existence of exact solutions for moving solitons does not imply the
integrability of the corresponding nonlinear lattice system. Nevertheless,
the availability of exact soliton solutions capable of moving with arbitrary
velocities across the lattice is a highly nontrivial feature, which permits
soliton collisions to occur. This is an interesting process which shall be
studied in Sec. 4.

To close this section we would like to notice that in the particular case
when $\rho =2\delta $ the Eq. (\ref{DDE2}) can be derived from a Hamiltonian
function (which is a dynamical invariant of the equation), provided that
suitable non-canonical Poisson brackets are introduced. As in this case ($%
\rho =2\delta $) no soliton solutions for Eq. (\ref{DDE2}) have been found,
we will not analyze this case here. However, the existence of a Hamiltonian
structure for Eq. (\ref{DDE2}) is an interesting issue which merit further
studies.

\section{Soliton embedding}

We have seen that the Eq. (\ref{DDE2}) has a continuous family of moving
lattice solitons. Now we aim to determine whether these solitons are
embedded in the spectrum of the radiation modes. To this end it is necessary
to check if the intrinsic frequency of the soliton (\ref{soliton}) is
contained (embedded) in the band of frequencies permitted to the
small-amplitude linear waves (phonon modes) which are able to propagate in
the lattice. If the soliton's frequency falls into this band, it will be
important to determine the wavenumbers corresponding to all the phonon modes
which have the same frequency as the soliton, since these modes will be
resonantly excited if the soliton is perturbed.

In order to determine the correct resonant wavenumbers it is necessary to
take into account that the solitons defined by Eqs. (\ref{soliton})-(\ref{q}%
) are \textit{moving solitons}. Due to this movement, the soliton's
intrinsic frequency must be compared to the linear spectrum of the phonon
waves \textit{in the reference frame which moves along with the soliton }%
(the transition to the moving reference frame, which is generally impossible
in a nonlinear lattice model, is always possible in the linearized equation
for the phonon modes). In this moving reference frame the coordinate $%
n\Delta x$ is replaced by $n\Delta x-at$, thus implying that the accordingly
defined frequency includes the Doppler shift. Additional arguments (based on
the Fourier expansion of the fields) showing that the resonance between
moving solitons and linear waves must be studied in the moving reference
frame were given in Refs. \cite{Espinosa} and \cite{Rodriguez2003} and, in a
more mathematically rigorous form, in Ref. \cite{Pelinovsky2005}.

According to the previous paragraph, we look for phonon modes in the form: $%
r_{n}\left( t\right) =\xi \,\exp \left[ i\Omega \,t-ik\left( n\Delta
x-at\right) \right] $, with $\xi $ being an arbitrary small amplitude. The
substitution of this expression in the linearized form of Eq. (\ref{DDE2})
leads to the dispersion relation: 
\begin{equation}
\Omega \left( k\right) =4\varepsilon _{0}\,\mathit{\sin }\left( k\Delta
x\right) \left[ 1-\mathit{\cos }\left( k\Delta x\right) \right] -ka,
\label{JFDR}
\end{equation}%
the last term being the above-mentioned Doppler shift. The moving soliton
will be embedded if its intrinsic frequency, which is $-\left( q+aQ\right) $
according to Eq. (\ref{soliton}), is contained in the range of values of the
function $-\Omega \left( k\right) $ with the real argument $k$ varying
within $-\infty <k<+\infty $. Since this range is the entire real axis (if $%
a\neq 0$), \emph{all} the moving lattice solitons given by Eq. (\ref{soliton}%
) are indeed embedded solitons.

In the particular case of $a=0$, the corresponding quiescent soliton will be
embedded if there is a real solution of the corresponding resonance
condition, which follows from equating expressions (\ref{q}) and (\ref{JFDR}%
): 
\begin{equation}
q=4\varepsilon _{0}\,\mathit{\sin }\left( k\Delta x\right) \left[ 1-\mathit{%
\cos }\left( k\Delta x\right) \right] .  \label{a=0}
\end{equation}%
It is easy to see that real solutions to Eq. (\ref{a=0}) exist for\ $%
\left\vert q\right\vert \leq 3\sqrt{3}|\varepsilon _{0}|$, otherwise the
static soliton will not be embedded.

We would like to observe that \textit{some }of the embedded solitons of the
family defined by the Eqs. (\ref{soliton})-(\ref{q}) move with velocities
which are contained within the range of phase velocities permitted to the
phonon modes. From (\ref{JFDR}) it follows that the phase velocities of
these modes (defined in the laboratory reference frame) are defined by: 
\begin{equation}
v\left( k\right) =\frac{4\,\varepsilon _{0}}{k\Delta x}\mathit{\sin }\left(
k\Delta x\right) \left[ 1-\mathit{\cos }\left( k\Delta x\right) \right] .
\label{V}
\end{equation}
Consequently, the soliton's velocity ($V=a/\Delta x$) will be within the
range of this function if the equation:

\begin{equation}
a=\frac{4\,\varepsilon _{0}}{k}\mathit{\sin }\left( k\Delta x\right) \left[
1-\mathit{\cos }\left( k\Delta x\right) \right] ,  \label{RC3}
\end{equation}
has real solution for $k$. Depending on the values of $\varepsilon _{0}$, $%
\rho $, $\delta $, $Q$ and $\Delta x$, real solutions to Eq. (\ref{RC3}) may
or may not exist.

From the previous paragraph it follows that the solitons of Eq. (\ref{DDE2})
can be classified in two groups. The first one contains those solitons whose
internal frequencies lie within the range of the function $-\Omega \left(
k\right) $, and whose velocities \textit{are not} contained in the range of
the function (\ref{V}). The second group includes those solitons which, in
addition of having wavenumbers within the range of $-\Omega \left( k\right) $%
, have velocities which are contained in the range of the function (\ref{V}%
). In the following, the solitons of the first group will be called \textit{%
single-embedded solitons}, and the solitons of the second group will be
referred to as \textit{double-embedded solitons}, to emphasize that in this
case both, the wavenumbers and the velocities of the solitons, are contained
in the corresponding linear spectra.

\section{Perturbed and colliding lattice solitons}

\subsection{Spectrum of emission from perturbed embedded solitons}

Since the intrinsic frequencies\ of the soliton solutions of Eq. (\ref{DDE2}%
) fall within the phonon spectrum of the system, one might expect that
resonances should occur between these solitons and the phonon modes, giving
rise to the emission of resonant radiation. However, the distinctive feature
of the ES is precisely the absence of this radiation. In the case of Eq. (%
\ref{cmKdV2}), the mechanism that suppresses the resonance between the ES
and the linear waves can be understood if one calculates the Fourier
transform\emph{\ }$\hat{u}\left( k,\omega \right) $\emph{\ }of the field%
\emph{\ }$u(x,t)\emph{.}$\emph{\ }As demonstrated in Ref. \cite%
{Rodriguez2003}, the calculation shows that $\hat{u}\left( k,\omega \right) 
\emph{\ }$is\ proportional to a quotient, $P_{2}\left( \omega \right)
/P_{1}\left( \omega \right) $, of two polynomials.\emph{\ }$P_{1}\left(
\omega \right) $\ is an image of the linear part of the equation, and
resonances occur at frequencies for which $P_{1}\left( \omega \right) $
vanishes, while $P_{2}\left( \omega \right) $\emph{\ }depends on the
nonlinear part of the equation. In order to calculate the Fourier transform
of the nonlinear term it is considered that $u\left( x,t\right) $ is a
single-humped function (such as a hyperbolic secant times a complex
exponential). In this way an explicit form for $P_{2}\left( \omega \right) $
can be obtained (see Eq. (52) in Ref. \cite{Rodriguez2003}). The crucial
point which explains why the ES do not radiate is that\emph{\ }$P_{2}\left(
\omega \right) $ vanishes at the same points where $P_{1}\left( \omega
\right) =0$,\emph{\ }provided that\emph{\ }$u(x,t)$\emph{\ }is \textit{%
exactly} an ES, and consequently the singularities in the expression $%
P_{2}\left( \omega \right) /P_{1}\left( \omega \right) $, which lead to the
resonances, cancel out. However, if $u(x,t)$\emph{\ }is an altered
(perturbed) ES, the exact cancelation does not take place, and hence \textit{%
transient emission} from the perturbed ES will take place at the frequencies
for which $P_{1}\left( \omega \right) =0$. Generally, if the soliton's
intrinsic frequency is\emph{\ }$\omega \equiv -\left( q+aQ\right) $ [as in
Eq. (\ref{soliton})],\emph{\ }and $\Omega \left( k\right) $ is\ the linear%
\emph{\ }dispersion relation for the phonon modes in the reference frame
moving along with the soliton, such as in Eq. (\ref{JFDR}), then\emph{\ }$%
P_{1}\left( \omega \right) =q+aQ-\Omega \left( k\right) $.\emph{\ }%
Consequently, a perturbed ES is expected to emit radiation at wavenumbers $%
-k $ determined by the resonance condition $\Omega \left( k\right) =q+aQ$.
In other words, the resonant wavenumbers are defined by the equation:

\begin{equation}
q+aQ=4\varepsilon _{0}\mathit{\sin }\left( k\Delta x\right) \left[ 1-\mathit{%
\cos }\left( k\Delta x\right) \right] -ka.  \label{r1}
\end{equation}%
To confirm that the perturbed solitons emit radiation at these wavenumbers,
let us analyze two examples. In the first case let us consider the following
parameters: $\varepsilon _{0}=1$, $\rho =-0.4$, $\delta =-0.85,$ $Q=0.61$
and $\Delta x=3$. In this case, the soliton (\ref{soliton}) is a
double-embedded one, with $A_{\mathrm{sol}}=0.56$, $w=12.12$, $a=7.75$ and $%
q=-5.1$. To verify that this soliton is double-embedded, in Fig. 1 we can
see the graph of the function $v\left( k\right) \Delta x$ [with $v\left(
k\right) $ given by Eq. (\ref{V})]. Since the maximum of this function is $%
7.967$, the soliton's parameter $a=7.75$ lies within the range of $v\left(
k\right) \Delta x,$ thus implying that the soliton is indeed
double-embedded. To perturb this soliton, we solved Eq. (\ref{DDE2})
numerically with the initial condition: 
\begin{equation}
r_{n}\left( t=0\right) =A_{0}\,\mathrm{sech}\left( \frac{n\Delta x}{w}%
\right) \,e^{i\,nQ\Delta x},  \label{ic1}
\end{equation}%
where the initial amplitude, $A_{0}=0.64$, is larger than the amplitude of
the exact soliton, $A_{\mathrm{sol}}=0.56$. According to Eq. (\ref{r1}), the
perturbed soliton will emit radiation at the wavenumbers: 
\begin{equation}
-\frac{k}{2\pi }=-0.114,\;-0.073,\ \ -0.008,\ +0.103,\ +0.129.  \label{wn1a}
\end{equation}

The numerical solution of the Eq. (\ref{DDE2}) with the initial condition (%
\ref{ic1}) is shown in Fig. 2 (for $t=20$), and the corresponding power
spectrum, $|r(k)|^{2}$, is displayed in Fig. 3 within the Brillouin zone, $%
-\pi /\Delta x<k<+\pi /\Delta x$, where the Fourier transform of the lattice
field is defined as:%
\begin{equation}
r(k)=\sum_{n}r_{n}\exp \left( ikn\Delta x\right) .  \label{FFT}
\end{equation}%
The spectrum shown in Fig. 3 contains a central component, which is
essentially the Fourier transform of the unperturbed soliton, and two
symmetric radiation bands, approximately located at $0.07\leq \left\vert
k/2\pi \right\vert \leq 0.13$. These radiation bands contain all the
expected resonant wavenumbers, except the value $-0.008$, which falls within
the central band. The \textit{symmetry} of the radiation bands is related to
the fact that the oscillating behavior (in time) of the soliton solutions of
Eq. (\ref{DDE2}) is described by a complex exponential $\mathit{\exp }\left(
i\omega _{0}t\right) $, whose real part oscillates as $\mathit{\cos }\left(
\pm \omega _{0}t\right) $. Due to this fact, a perturbed soliton is able to
excite the phonon modes with frequencies $\pm \omega _{0}$, thus accounting
for the occurrence of symmetrically reflected resonance numbers.

As a second example, let us perturb the single-embedded soliton
corresponding to the parameters: $\varepsilon _{0}=-4.8$, $\rho =4.8$, $%
\delta =10.3,$ $Q=7.9$ and $\Delta x=0.4$. In this case the parameters of
the soliton are: $A_{\mathrm{sol}}=0.381$, $w=1.503$, $a=15.812$ and $%
q=-0.771$. To confirm that this soliton is single-embedded, in Fig. 4 we
display the phase velocity (times $\Delta x$) of the radiation waves as
given by Eq. (\ref{V}). From this figure it is evident that the soliton's
parameter $a=15.812$ lies outside the range of the function $v\left(
k\right) \Delta x$.

The resonance condition (\ref{r1}) defines in this case only one resonant
wavenumber located at $-k/2\pi =-1.249$, which is very close to one of the
edges of the Brillouin zone $\left\vert k/2\pi \right\vert \leq 1/\left(
2\Delta x\right) =1.25$. As in the previous example, we also expect a
symmetrically reflected radiation peak located at $-k/2\pi =+1.249$, due to
a complex factor describing the oscillating behavior of the perturbed
soliton.

The numerical solution of Eq. (\ref{DDE2}) corresponding to an initial
condition of the form (\ref{ic1}) with $A_{0}=0.35$ (which is smaller than
the amplitude of the exact soliton) shows that the perturbed pulse emits a
radiation wavetrain to the left, as seen in Fig. 5. The spectrum of the
numerical solution, shown in Fig. 6, exhibits two symmetric radiation peaks
at $-k/2\pi =\pm 1.23$, which are very close to the expected values.

\subsection{Robustness of the embedded solitons}

The emission of radiation from a perturbed linearly stable ELS may result in
its full decay, and consequently its stability (or, better said, robustness)
must be tested in long simulations. It is relevant to mention that in the
continuum limit, the ES solutions of Eq. (\ref{cmKdV2}) are stable because
this equation gives rise to a continuous family of ES with arbitrary energy,
and hence a perturbed soliton can shed off a part of its energy and settle
down to an ES solution with a smaller energy \cite{Rodriguez2003}. In
continuum models supporting isolated ES, the perturbed pulse may not be able
to find a stationary soliton with a lower energy toward which it could
relax, and hence it may go on radiating till complete decay. This scenario,
referred to as the \textquotedblleft semi-stability" (alias nonlinear
instability) of the ES, really occurs in the continuum model with the
combined $\chi ^{(2)}:\chi ^{(3)}$ nonlinearity \cite{Yang99,PhysD}.

In the present case, the stability of the ELS may be secured if a perturbed
soliton can find a way to relax, after shedding off some radiation, to a new
stationary soliton shape. Within the bounds of the exact solution family
given by Eqs. (\ref{soliton})-(\ref{q}), this process does not seem
plausible, as all the exact solutions have a unique value of the amplitude.
Nevertheless, it may happen that the exact family is only a subset of a more
general manifold of soliton solutions. The possible existence of such an
extended soliton family would not only provide for the stability as
explained above, but it is a fundamentally important issue by itself.

To clarify the eventual fate of the perturbed solitons of Eq. (\ref{DDE2}),
we now display some typical examples. In the first place, let us consider
the double-embedded soliton corresponding to $\varepsilon _{0}=1$, $\rho
=-0.4$, $\delta =-0.85$, $Q=0.61$ and $\Delta x=3$. As stated before, the
shape of this soliton is defined by the parameters $A_{\mathrm{sol}}=0.56$, $%
w=12.12$\emph{\ }and\emph{\ }$a=7.75$. If this soliton is perturbed by
increasing its amplitude [for instance, taking the initial condition (\ref%
{ic1}) with $A_{0}=0.64$], the perturbed pulse starts to oscillate. In the
course of the evolution, the oscillations gradually fade away due to the
radiation loss, as shown by the upper curve in Fig. 7, and the perturbed
soliton tends to settle down into a stationary shape, which is quite
possible, as the initial perturbation increased its norm.

On the other hand, if the height of the initial pulse is \emph{smaller} than
the amplitude of the exact soliton (for example, $A_{0}=0.48$), the norm of
the perturbed pulse becomes smaller than the exact soliton's norm, and
therefore the pulse will not be able to relax back to the exact soliton. As
explained above, such a perturbation might lead to the destruction of the
soliton, unless it can find another stationary shape to which it may relax
to. The evolution of this perturbed soliton was numerically simulated, and
the lower curve in Fig. 7 shows the behavior of the soliton's amplitude. As
we can see in this figure, the perturbed soliton does not decay
indefinitely. It again exhibits a damped oscillatory behavior (although the
oscillation period is much longer than in the previous case, when the
perturbation increased the norm), and it finally approaches a new
equilibrium state. The behavior exhibited in Fig.7 indicates that the
double-embedded lattice solitons are \emph{completely stable} pulses and
also, as conjectured above, that the exact solitons seem to be just a subset
of a broader family of discrete solitons. The complete characterization of
this broader family lies outside the scope of the present work and it will
not be addressed here.

The stability of the single-embedded lattice solitons of Eq. (\ref{DDE2}) is
even clearer. Let us consider, for instance, the soliton corresponding to $%
\varepsilon _{0}=-4.8$, $\rho =4.8$, $\delta =10.3$, $Q=7.9$, and $\Delta
x=0.4$. In this case, $A_{\mathrm{sol}}=0.381$, $w=1.503$,\emph{\ }$q=-0.771$%
\emph{\ }and\emph{\ }$a=15.812$. As the value of the parameter $a$ lies
outside the range of the function $v(k)\Delta x$ (see Fig. 4), the soliton
is indeed single-embedded (\textit{i.e.},only in terms of its frequency, but
not according to\ its velocity). To perturb this soliton, we first took the
initial condition (\ref{ic1}) with an increased amplitude, $A_{0}=0.415$.
This perturbed pulse stabilizes itself quite rapidly, as we can see from the
time evolution of its height, which is presented in the upper curve of Fig.
8. In a similar way, if the initial amplitude is \emph{decreased} to $%
A_{0}=0.345$ (recall that the amplitude of the exact soliton is $A_{\mathrm{%
sol}}=0.381$), the pulse also relaxes quickly, its height evolving as shown
by the lower curve in Fig. 8. This figure indicates that the single-embedded
soliton of Eq. (\ref{DDE2}) tested in these simulations is a\emph{\ }stable
solution. Figure 8 also shows that when the initial pulse is higher (lower)
than the exact soliton, the amplitude of the established steady-state pulse
will be still larger (smaller) than the amplitude $A_{\mathrm{sol}}$ of the
exact soliton solution. A similar behavior was observed in the case of the
ES of Eq. (\ref{cmKdV2}), and an explanation of this behavior (by means of a
variational analysis) was provided in Ref. \cite{Rodriguez2003}.

A comparison of Figs. 7 and 8 reveals an interesting difference: while the
amplitude of the perturbed double-embedded soliton shown in Fig. 7 exhibits
well-defined oscillations, the amplitude of the single-embedded soliton seen
in Fig. 8 evolves almost monotonically. This qualitative difference may be
related to the fact that the resonance condition (\ref{r1}) determines 
\textit{five} resonant wavenumbers in the case studied in Fig. 7, and 
\textit{only one} in the case corresponding to Fig. 8. Actually, a precise
explanation of the different behaviors exhibited in Figs. 7 and 8 remains a
challenge for future work.

We again stress that the perturbed lattice solitons of Eq. (\ref{DDE2})
stabilize as new stable pulses whose heights are \emph{different} from the
unique value of the amplitude of the exact soliton solutions given by Eq. (%
\ref{A}). As said above, this observation strongly suggests that the exact
solutions given by Eqs. (\ref{soliton})-(\ref{q}) constitute just a
particular subset of a \emph{wider} soliton family of the Eq. (\ref{DDE2}),
which is a subject for a separate work.

Our simulations, performed at many values of the parameters, have \emph{never%
} revealed an unstable soliton, even if the initial perturbation was
actually large. An additional example is displayed in Fig. 9, for $%
\varepsilon _{0}=-1$, $\rho =0.4$, $\delta =0.85$, $\Delta x=3$ and $Q=0.61$%
. The exact soliton corresponding to these values has $A_{\mathrm{sol}}=0.56$%
, $w=12.12$, $a=7.75$, and $q=-5.103$, while the initial condition was taken
in the form of Eq. (\ref{ic1}) with $A_{0}=0.48$, which is essentially lower
than the amplitude of the exact solution. In this case again, the amplitude
of this perturbed pulse exhibits damped oscillations and a very clear trend
to the stabilization into a new stationary soliton, which does not belong to
the family of the exact solutions.

\subsection{Collisions between embedded solitons}

Once knowing that the ELS of Eq. (\ref{DDE2}) are stable against
perturbations, it is natural to consider their robustness against
collisions. To this end, we simulated collisions between two ELS moving in
opposite directions. A typical example can be displayed for $\varepsilon
_{0}=-4.8$, $\rho =4.8$, $\delta =10.3$ and $\Delta x=0.4$. In this case,
all the solitons belonging to family (\ref{soliton})-(\ref{q}) have the same
amplitude and width, $A_{\mathrm{sol}}=0.38$ and $w=1.5$, but may differ in
the velocity, which is controlled by the parameter $Q$. We can take a
soliton moving to right, with $Q_{1}=3.315$ and $a_{1}=15.7$, and a second
one travelling to left, with $Q_{2}=7.9$ and $a_{2}=-9.0$. Initially, these
solitons are placed at $n\Delta x=-20$ and $n\Delta x=12$, respectively.

The collision between these solitons is shown in Fig. 10, which demonstrates
that they recover their shapes and velocities after the collision. In Fig.
11, we specially display the shape of the solitons after the collision (at $%
t=11$). This figure shows that no radiation is emitted by the solitons after
the collision, and it confirms that the solitons keep the same value of the
amplitude, $A_{\mathrm{sol}}=0.38$, which they had before the collision. At
the moment shown ($t=11$), the centers of the pulses are located at $n\Delta
x=-78.6$ and $n\Delta x=139.2$. Without the collision, they would be found
(at the same moment of time) at the points $n\Delta x=-87\ $and $n\Delta
x=152.7$, respectively. We thus conclude that the only effect of the
collision is a shift in the position of the solitons (as occurs usually in
integrable systems). The same result was observed in simulations performed
at other values of the parameters.

\section{Summary and conclusions}

In this work we have introduced a new one-dimensional dynamical lattice
model. The underlying equation (\ref{DDE2}) includes next-nearest-neighbor
couplings, and it may be regarded as a discretization of the complex
modified KdV equation, or as an extension of the Ablowitz-Ladik model (in
the staggered form). The model may apply to the description of a
Bose-Einstein condensate with dipole interactions between atoms, trapped in
a deep optical lattice.

A family of exact lattice-soliton solutions of the discrete equation was
found, with the following characteristics:

\noindent (i) These solitons can move with arbitrary constant velocities
(within a certain interval), without being hindered by the lattice
discreteness. On the other hand, all the solitons have the same amplitude.

\noindent (ii) All the soliton solutions are indeed embedded (with the
exception of a part of the subfamily of zero-velocity solitons) because
their internal frequency lies within the phonon band (calculated in a moving
reference frame which moves along with the soliton, and contains the
corresponding Doppler shift).

\noindent (iii) The embedded solitons of Eq. (\ref{DDE2}) are not isolated
solutions (unlike most other models admitting embedded solitons), but form a
continuous two-parameter family.

\noindent (iv) Numerical simulations show that all the solitons are stable.

\noindent (v) The exact soliton solutions with the fixed amplitude
constitute a subset a larger family of stable discrete moving solitons,
which can be found in a numerical form. In particular, under a perturbation
that reduces the soliton's norm, a perturbed soliton relaxes to a new one
which belongs to the extended family.

\noindent (vi) Collisions between the moving solitons known in the exact
form are completely elastic (according to the results of systematic
simulations), leading solely to a shift of the solitons' positions.

Some of these features, such as the existence of a continuous family of
moving lattice solitons and the elasticity of the collisions between them
(as confirmed up to the numerical accuracy), suggest that the model may have
a chance to be integrable. Investigation of this possibility poses a
challenge for further studies of the model.

\section*{Acknowledgements}

We sincerely thank A. Minzoni for his interest in this project and J. Yang
for his useful comments and for providing us with a preprint of Ref. \cite%
{Dmitry}. DGSCA-UNAM (\textit{Direcci\'{o}n General de Servicios de C\'{o}%
mputo Acad\'{e}mico} of Universidad Nacional Aut\'{o}noma de M\'{e}xico) is
acknowledged for granting us access to their computers Bakliz and Berenice.
This work was financially supported from the grant DGAPA-UNAM IN112503. One
of the authors (BAM) appreciates financial support from FENOMEC during his
stay in Mexico.

\pagebreak

\begin{center}
\bigskip \textbf{Figure captions}
\end{center}

Fig. 1. The phase velocity of the linear waves (times $\Delta x$) as a
function of $k$ [i.e\textit{.,} the right-hand side of Eq. (\ref{RC3})] for $%
\varepsilon _{0}=1$ and $\Delta x=3$.\medskip 

Fig. 2. The perturbed double-embedded lattice soliton, at $t=20$, which
evolved from the initial condition (\ref{ic1}), with $A_{0}=0.64$, $w=12.12$%
, $Q=0.61$, and $\Delta x=3$. The parameters of Eq. (\ref{DDE2}) are $%
\varepsilon _{0}=1$, $\rho =-0.4$, and $\delta =-0.85$. In this figure and
below, the shape of the soliton is shown as $|r_{n}|$ versus $n$.\medskip 

Fig. 3. The power spectrum of the numerical solution corresponding to the
perturbed double-embedded lattice soliton shown in Fig. 2.\medskip 

Fig. 4. The phase velocity of the linear waves (times $\Delta x$) as a
function of $k$, for $\varepsilon _{0}=-4.8$ and $\Delta x=0.4$.\medskip 

Fig. 5. The perturbed single-embedded lattice soliton is shown at $t=10$. It
has evolved from the initial condition (\ref{ic1}), with $A_{0}=0.35$, $%
w=1.503$, $Q=7.9$, and $\Delta x=0.4$. The parameters of Eq. (\ref{DDE2})
are $\varepsilon _{0}=-4.8$, $\rho =4.8$, and $\delta =10.3$. The velocity
of the soliton is defined by the parameter $a=15.812$.\medskip 

Fig. 6. The power spectrum of the numerical solution corresponding to the
perturbed single-embedded lattice soliton shown in Fig. 5.\medskip 

Fig. 7. Evolution of the amplitudes of two perturbed double-embedded lattice
solitons. In both cases, the initial conditions are given by Eq. (\ref{ic1}%
), with $w=12.12$, $Q=0.61$, and $\Delta x=3$. The parameters of Eq. (\ref%
{DDE2}) are $\varepsilon _{0}=1$, $\rho =-0.4$, and $\delta =-0.85$. The
upper and lower curves correspond, respectively, to the initial amplitudes $%
A_{0}=0.64$ and $A_{0}=0.48$, which are larger and smaller than the value $%
A_{\mathrm{sol}}=0.56$ for the exact soliton solution in this case.\medskip 

Fig. 8. Evolution of amplitudes of two perturbed single-embedded lattice
solitons. In both cases, the initial conditions are taken as per Eq. (\ref%
{ic1}), with $w=1.5$, $Q=7.9$, and $\Delta x=0.4$. Parameters of Eq. (\ref%
{DDE2}) are $\varepsilon _{0}=-4.8$, $\rho =4.8$, and $\delta =10.3$. The
upper and lower curves correspond, respectively, to the initial amplitudes $%
A_{0}=0.415$ and $A_{0}=0.345$, which are larger and smaller that the value $%
A_{\mathrm{sol}}=0.38$ corresponding to the exact soliton in this
case.\medskip 

Fig. 9. Evolution of the amplitude of a perturbed soliton corresponding to
the parameters $\varepsilon _{o}=-1$, $\rho =0.4$, $\delta =0.85$ and $%
\Delta x=3$. The initial pulse is take in the form of Eq. (\ref{ic1}) with $%
A_{0}=0.48$, $r=0.61$ and $w=12.12$.\medskip 

Fig. 10. Collision of two single-embedded lattice solitons, at $\varepsilon
_{0}=-4.8$, $\rho =4.8$, $\delta =10.3$ and $\Delta x=0.4$. The solitons
moving to the right and left are determined by parameters, respectively, $%
Q_{1}=3.315$, $a_{1}=15.7$, and $Q_{2}=7.9$, $a_{2}=-9.0$.\medskip 

Fig. 11.The shape of the lattice field after the collision shown in Fig.10,
at $t=11$.

\pagebreak 

\begin{center}
\bigskip 

\FRAME{ftbpF}{250.9375pt}{190.5pt}{0pt}{}{}{Figure}{\special{language
"Scientific Word";type "GRAPHIC";maintain-aspect-ratio TRUE;display
"USEDEF";valid_file "T";width 250.9375pt;height 190.5pt;depth
0pt;original-width 310.1875pt;original-height 234.875pt;cropleft "0";croptop
"1";cropright "1";cropbottom "0";tempfilename
'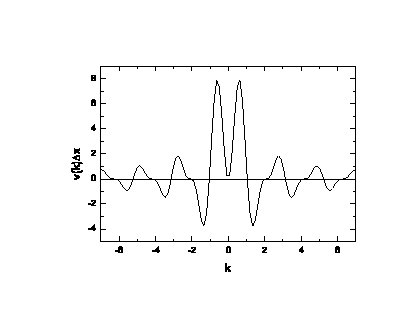';tempfile-properties "XPR";}}
\end{center}

\FRAME{ftbpF}{291.125pt}{229.875pt}{0pt}{}{}{Figure}{\special{language
"Scientific Word";type "GRAPHIC";maintain-aspect-ratio TRUE;display
"USEDEF";valid_file "T";width 291.125pt;height 229.875pt;depth
0pt;original-width 287.5625pt;original-height 226.625pt;cropleft "0";croptop
"1";cropright "1";cropbottom "0";tempfilename
'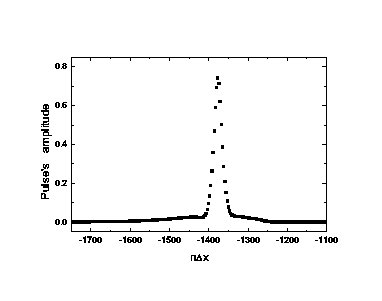';tempfile-properties "XPR";}}

\FRAME{ftbpF}{298.6875pt}{229.875pt}{0pt}{}{}{Figure}{\special{language
"Scientific Word";type "GRAPHIC";maintain-aspect-ratio TRUE;display
"USEDEF";valid_file "T";width 298.6875pt;height 229.875pt;depth
0pt;original-width 295.125pt;original-height 226.625pt;cropleft "0";croptop
"1";cropright "1";cropbottom "0";tempfilename
'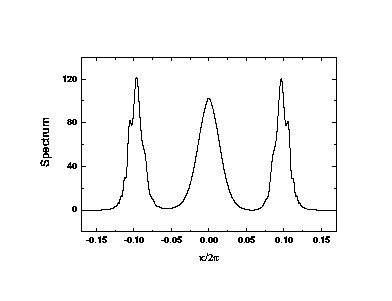';tempfile-properties "XPR";}}

\FRAME{ftbpF}{310.0625pt}{236.625pt}{0pt}{}{}{Figure}{\special{language
"Scientific Word";type "GRAPHIC";maintain-aspect-ratio TRUE;display
"USEDEF";valid_file "T";width 310.0625pt;height 236.625pt;depth
0pt;original-width 306.375pt;original-height 233.375pt;cropleft "0";croptop
"1";cropright "1";cropbottom "0";tempfilename
'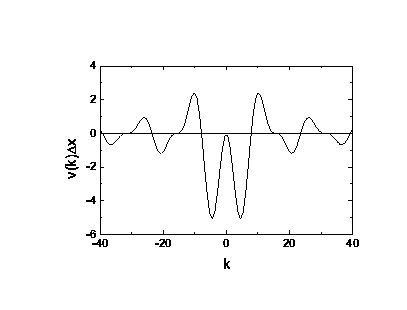';tempfile-properties "XPR";}}

\FRAME{ftbpF}{296.4375pt}{229.875pt}{0pt}{}{}{Figure}{\special{language
"Scientific Word";type "GRAPHIC";maintain-aspect-ratio TRUE;display
"USEDEF";valid_file "T";width 296.4375pt;height 229.875pt;depth
0pt;original-width 292.875pt;original-height 226.625pt;cropleft "0";croptop
"1";cropright "1";cropbottom "0";tempfilename
'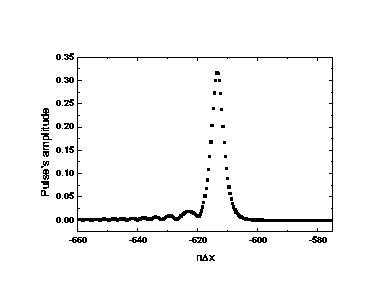';tempfile-properties "XPR";}}

\FRAME{ftbpF}{296.4375pt}{231.375pt}{0pt}{}{}{Figure}{\special{language
"Scientific Word";type "GRAPHIC";maintain-aspect-ratio TRUE;display
"USEDEF";valid_file "T";width 296.4375pt;height 231.375pt;depth
0pt;original-width 292.875pt;original-height 228.125pt;cropleft "0";croptop
"1";cropright "1";cropbottom "0";tempfilename
'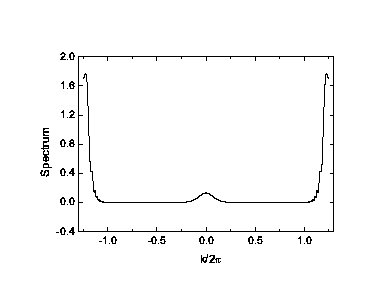';tempfile-properties "XPR";}}

\FRAME{ftbpF}{297.1875pt}{229.875pt}{0pt}{}{}{Figure}{\special{language
"Scientific Word";type "GRAPHIC";maintain-aspect-ratio TRUE;display
"USEDEF";valid_file "T";width 297.1875pt;height 229.875pt;depth
0pt;original-width 293.625pt;original-height 226.625pt;cropleft "0";croptop
"1";cropright "1";cropbottom "0";tempfilename
'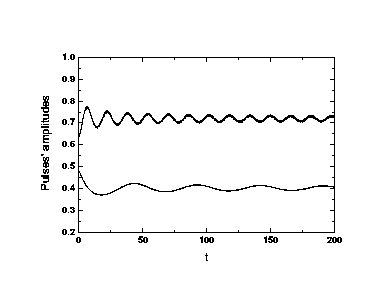';tempfile-properties "XPR";}}

\FRAME{ftbpF}{227.4375pt}{190.5pt}{0pt}{}{}{Figure}{\special{language
"Scientific Word";type "GRAPHIC";maintain-aspect-ratio TRUE;display
"USEDEF";valid_file "T";width 227.4375pt;height 190.5pt;depth
0pt;original-width 294.375pt;original-height 246.1875pt;cropleft "0";croptop
"1";cropright "1";cropbottom "0";tempfilename
'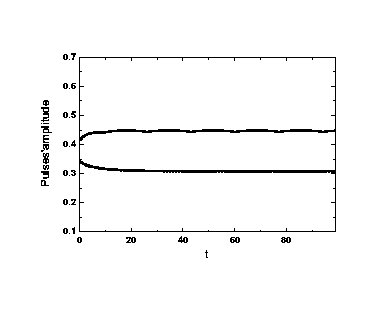';tempfile-properties "XPR";}}

\FRAME{ftbpF}{307pt}{235.875pt}{0pt}{}{}{Figure}{\special{language
"Scientific Word";type "GRAPHIC";maintain-aspect-ratio TRUE;display
"USEDEF";valid_file "T";width 307pt;height 235.875pt;depth
0pt;original-width 303.375pt;original-height 232.625pt;cropleft "0";croptop
"1";cropright "1";cropbottom "0";tempfilename
'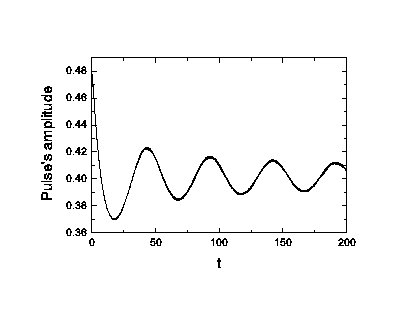';tempfile-properties "XPR";}}

\FRAME{ftbpF}{240.125pt}{190.5625pt}{0pt}{}{}{Figure}{\special{language
"Scientific Word";type "GRAPHIC";maintain-aspect-ratio TRUE;display
"USEDEF";valid_file "T";width 240.125pt;height 190.5625pt;depth
0pt;original-width 307.875pt;original-height 243.9375pt;cropleft "0";croptop
"1";cropright "1";cropbottom "0";tempfilename
'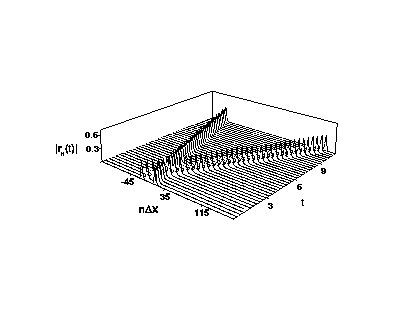';tempfile-properties "XPR";}}

\FRAME{ftbpF}{301.75pt}{232.125pt}{0pt}{}{}{Figure}{\special{language
"Scientific Word";type "GRAPHIC";maintain-aspect-ratio TRUE;display
"USEDEF";valid_file "T";width 301.75pt;height 232.125pt;depth
0pt;original-width 298.125pt;original-height 228.875pt;cropleft "0";croptop
"1";cropright "1";cropbottom "0";tempfilename
'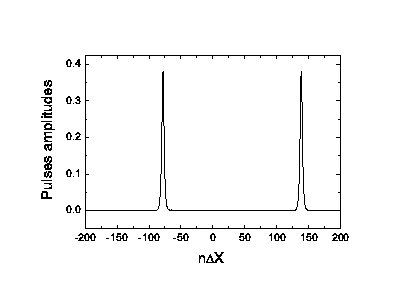';tempfile-properties "XPR";}}

\end{document}